\documentclass[
prmaterials,
reprint,               
superscriptaddress,     
floatfix
]{revtex4-2}

\usepackage[T1]{fontenc}
\usepackage[utf8]{inputenc}
\usepackage{lmodern}
\usepackage{amsmath,amssymb,amsfonts}
\usepackage{graphicx}
\usepackage{xcolor}
\usepackage{microtype}
\usepackage{hyperref}
\hypersetup{
	colorlinks=true,
	linkcolor=blue,
	citecolor=blue,
	urlcolor=blue
}

\begin{document}
	
	\title{Efficient molecular dynamics simulation of 2D penta-silicene materials \\ using machine learning potentials}

	\author{Le Huu Nghia}
	\affiliation{College of Natural Sciences, Can Tho University, 3-2 Road, Can Tho City 94000, Vietnam} 
	
	\author{Pham Thi Bich Thao}
	\affiliation{College of Natural Sciences, Can Tho University, 3-2 Road, Can Tho City 94000, Vietnam}
	
	\author{Truong Do Anh Kha}
	\affiliation{College of Natural Sciences, Can Tho University, 3-2 Road, Can Tho City 94000, Vietnam}

	\author{Vo Khuong Dien}
	\affiliation{College of Natural Sciences, Can Tho University, 3-2 Road, Can Tho City 94000, Vietnam}
	
	\author{Nguyen Thanh Tien}
	\email{nttien@ctu.edu.vn}
	\affiliation{College of Natural Sciences, Can Tho University, 3-2 Road, Can Tho City 94000, Vietnam}
	
	\date{\today}

\begin{abstract}
	Machine Learning Interatomic Potentials (MLIPs) are a modern computational method that allows achieving near-quantum mechanical accuracy (DFT) while still describing large-scale systems in molecular dynamics (MD) simulations. In this work, we use MLIP from DeepMD package and the classical Tersoff potential for SiC (Tersoff.SiC potential) to fully and accurately describe atomic interactions and apply them to molecular dynamics simulations of penta silicene sheet. The results show that the melting points (T$_g$) temperatures of the system in the canonical NVT and isobaric NPT sets are 632 K and 606 K, while the Tersoff.SiC potential have the high melting points, respectively. In addition, the radial distribution function exhibits characteristic peaks at interatomic distances of 2.275 \AA \text{} and 2.375 \AA, while the Tersoff.SiC potential only describe distance of 2.375 \AA. Furthermore, penta silicene was also simulated using on-the-fly machine learning for 10 ps to evaluate the structural stability of the system. This study investigates the thermodynamic properties of two-dimensional penta silicene sheets with pentagonal structures using a high-precision, cost-effective method, contributing further evidence to support experimental synthesis and opening up potential future applications of this material.
	
\end{abstract}

\maketitle

\section{Introduction}\label{sec:intro}
Two-dimensional (2D) materials are a groundbreaking group of materials that have received much research attention \cite{SHEN20221}, including pentagonal  of carbon, silicon, and germanium, ... The first idea of Shunjong Zhang, et al was to find a 2D carbon allotrope structure "like Cairo pentagonal tiles", and named it penta graphene \cite{Shunhong, https://doi.org/10.1002/smll.201600382,BEKHTISIAD2018870}. Similar to penta graphene, penta silicene compare between sp²–sp³ hybridization of silicon. However, according to first-principle calculations, this perfect penta silicene structure is dynamically unstable and only stabilized by hydrogen-induced \cite{C5TC02504D}, the type of stacking mode AA or AB \cite{C6CP03200A}. Yaguang Guo, et al \cite {PhysRevApplied.11.064063} (2019) reported the model penta silcene dynamically stabilized through first-principle calculations, by moving the Si$_3$ on the dimer, leading to a break in symmetry, generate the tilting and reduce the strong Coulomb interactions between the charged Si atoms. The existence was confirmed by Jorge I. Cerda, et al \cite{doi.org/10.1038/ncomms13076} who successfully synthesized single-strand and double-strand Si pentagonal nano-ribbons on a silver {Ag(110)} surface \cite{PhysRevLett.117.276102,doi:10.1021/acs.nanolett.8b00289}.
\\Recent developments in machine learning models have shown an increasingly efficient and accurate machine learning interatomic potential (MLIP) representation solution \cite{doi:10.1021/acs.jpcc.5c03470, 10.1063/5.0251741, Novikov_2021}. The DeepMD-kit package \cite{WANG2018178, 10.1063/5.0155600, doi:10.1021/acs.jctc.2c00102} is a machine learning model that constructs force fields from Density Functional Theory (DFT) data using deep machine learning algorithms. The DeepMD-kit package serves as a bridge between DFT and Molecular Dynamics (MD) simulations. The ab initio-trained datasets allow large-scale system simulations on the LAMMPS package \cite{THOMPSON2022108171} while maintaining quantum mechanical accuracy. Yaguang Guo et al \cite {PhysRevApplied.11.064063} proposed the model penta silcene dynamically stabilized and also performed the Monte Carlo simulation with 50×50 supercell. They reported the ferroelectricity with the high Curie temperature of penta silicene Tc=1190K. However, the phase transitions and MD simulations with ab initio-trained datasets in LAMMPS package \cite{THOMPSON2022108171} have not been studied yet. 
\\The cost of ab-initio MD simulations is computationally expensive, because electrons are treated entirely quantum mechanically in DFT \cite{PhysRevLett.122.225701, Vandermause2020}. Additionally, the phonon dispersion also requires the estimation of the second interatomic force constants (IFCs) \cite{10.1063/5.0251741, MORTAZAVI2021107583}, which are usually obtained by DFT calculations on the supercell lattice. The above calculation is also expensive and time-consuming, because requiring single-point from DFT force calculations \cite{MORTAZAVI2021107583}. Machine Learning Force Field (MLFF) is a modern solution that builds computationally computation between ab-initio MD simulations and machine learning to construct interatomic potentials from ab-initio MD data \cite{ZHANG2023158141, 10.1063/5.0251741}. In this approach, electrons are treated quantum mechanically using DFT, which results in a lower computational cost. Instead of using from experimental potential, the MLFF approach on-the-fly training of potentials, based on the process of predicting energy, force, stress tensor, and their uncertainty on a given structure using existing force fields \cite{PhysRevB.100.014105}. 
\\In this study, we propose two computational strategies based on machine-learning interatomic potentials (MLIPs). For evaluating the thermal properties of monolayer penta silicene within MD simulation, we employ an MLIP from ab initio trained dataset using the DeepMD-kit package, enabling large-scale simulations while preserving quantum-mechanical accuracy. In parallel, we also performed simulations on the Tersoff.SiC potential to compare the results. For checking thermal stability, we performed instantaneous machine learning calculations with ab initio molecular dynamics (AIMD) simulations and evaluated the thermal stability of the structure.

\section{Computational details}
All first-principles calculations were implemented using the Vienna Ab initio Simulation Package (VASP) \cite{PhysRevB.54.11169, https://doi.org/10.1002/jcc.21057}. The projector augmented wave pseudopotentials with the Perdew–Burke –Ernzerhof (PBE) exchange–correlation functional \cite{PhysRevLett.77.3865} and the projector augmented wave (PAW) \cite{PhysRevB.59.1758, PhysRevB.50.17953} method were used to treat exchange-correlation effects. The cut-off energy was set at 400 eV. The convergence criterion of 10$^{-2}$  eV/$\AA$ and 10$^{-6}$ eV was used for the force and energy in the calculations, respectively. A vacuum of 18 \AA \text{} was added along the z-direction of the structure. A Monkhorst-Pack grid of 12x12x1 was used to sample the Brillouin zone \cite{PhysRevB.13.5188}. The phonon dispersion are calculated by PHONOPY \cite{PhysRevB.78.134106} code using finite displacement method \cite{PhysRevLett.78.4063}. A 4x4x1 supercell is constructed, and a 3x3x1 k-point mesh was used to calculate the atomic forces, with high accurary energy convergence criteria 10$^{-8}$ eV.   
\subsection{Machine Learning Interatomic Potential}
\begin{figure*}[h]
	\centering
	\includegraphics[scale=0.9]{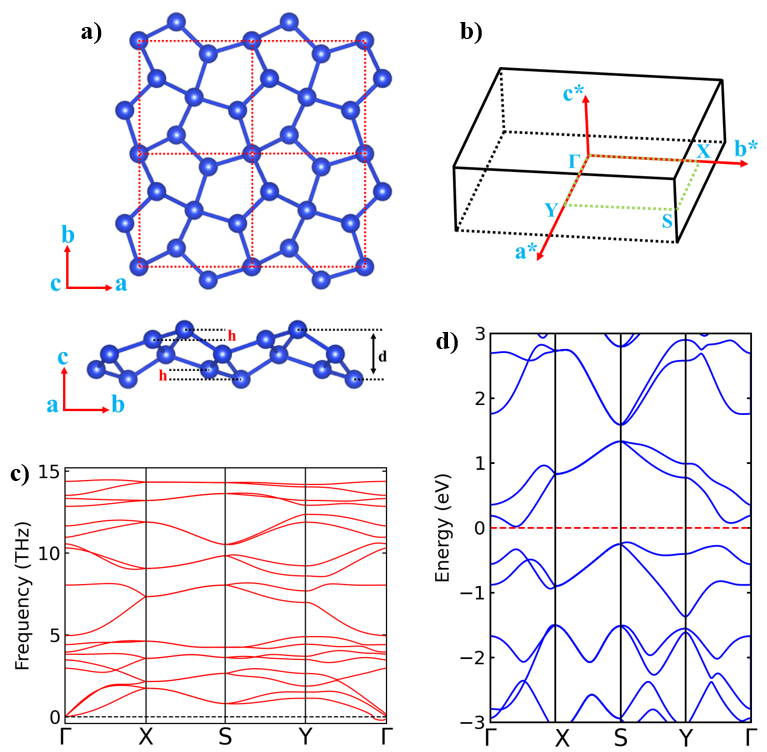}
	\caption{a) The optimized structure of penta silicene monolayer is shown in both top and side view. b) The first Brillouin zone. c) The phonon dispersion of the penta silicene monolayer is presented based on DFT calculations. d) Electronic band structure calculated using the PBE functional.}
	\label{fig1}
\end{figure*}
The workflow employed for the development data of penta silicene machine learning interatomic potential (MLIP). We carried out performing AIMD simulations on a 4x4x1 supercell with 96 atoms at temperatures from 100 K to 900 K with a step of 100 K. The time steps used 0.5 fs, and the number for steps was only 1000 time steps for each temperature model. Only the $\Gamma$-point was used for k-space sampling in all calculations. The training data for each of these models were generated from canonical ensemble (NVT) simulations. The temperature control is achieved by the Nose-Hoover thermostat \cite{10.1063/1.449071, PhysRevA.31.1695}. 

The MLIP was performed using the DeepMD-kit package \cite{WANG2018178} based on data of forces, energies, virial stresses from the AIMD calculations to generate deep learning potential. We consider the $N$ atoms of the structure, and the total energy $E$ given by \cite{WANG2018178, D4CP00997E}:

\begin{equation}
	E = \sum_i E_i, 
\end{equation}
with $i$ is the indexes of the atoms, and $E_i$ is each atomic energy, and we determined by the position of the $i{th}$ atom and its near neighbors, and the energy $E_i$ given by \cite{WANG2018178, D4CP00997E}:
\begin{equation}
	E_i = E_{s(i)}(\mathbf{R}_i, \{\mathbf{R}_j \,|\, j \in N_{R_c}(i)\}),
\end{equation}
where $R_C$, $N_{Rc}$, and $s(i)$ represent the cut-off radius, the set of the neighbours of atom $i$ within the cut-off radius and chemical species of the $ith$ atom, respectively. 

DeepMD-kit package is calculated by performing the descriptors ($D_{ij}$) to generate an atomic chemical environment into system, where the descriptor matrix $D_{ij}$ is built for atom $j$ ensuring that the atom $j$ is determined within the designated cutoff radius relative to the atom $i$ \cite{WANG2018178, D4CP00997E}.
\begin{equation}
	D_{ij} = 
	\left\{
	\frac{1}{R_{ij}},
	\frac{x_{ij}}{R_{ij}},
	\frac{y_{ij}}{R_{ij}},
	\frac{z_{ij}}{R_{ij}}
	\right\}
\end{equation}
where $x_{ij}$, $y_{ij}$, and $z_{ij}$ are the coordinates, respectively, and $R_{ij}$ represents the distance between $i$ and $j$. And then, we write the each atomic energy by function \cite{WANG2018178, D4CP00997E}:
\begin{equation}
	E_i = \lambda_i (D_{ij})
\end{equation}
where $\lambda$ represents the multilayer perceptron with the hidden layer. 

The  loss function (L) is given by \cite{WANG2018178, D4CP00997E}:
\begin{equation}
	L(p_\epsilon, p_f, p_\xi) = 
	\frac{p_\epsilon}{N} \Delta E^2 
	+ \frac{p_f}{3N} \sum_i |\Delta \textbf{F}_i|^2 
	+ \frac{p_\xi}{9N} \sum_i \|\Delta v_i\|^2
\end{equation}
where $\Delta E$, $\Delta \textbf{F}_i$, and $\Delta v$ are denote root mean square  error (RMSE) in energy, force, and virial, respectively. The prefactors $p_\epsilon$, $p_f$, and $p_\xi$ are adjustable coefficients. $N$ represents the total number of atoms, and $\left|{..}\right|$, and  $\left\|{..}\right\|$ represent the norm of the force vector and virial tensor \cite{D4CP00997E}. 

The following training parameters were set up within models: the descriptor “se\_a” was used the cut-off radius and smoothing radius were \textit{rcut} = 6 \text{\AA} and \textit{rcut\_smth} = 5.8 \text{\AA}, respectively. There hidden layers contained 25, 50, and 100 neurons and the fitting network consisted of 240 neurons each. The start learning rate and the end learning rate were \textit{start\_lr} = 10$^{-3}$, and \textit{stop\_lr} = 3.51e$^{-8}$, and the number of training batch stop was \textit{stop\_batch} = 10$^{6}$ to improve accuracy and avoid over-fitting of the training data during model training, respectively. Once the MLIP was generated, we conducted classical molecular dynamics simulations from this potential.

\subsection{MLIP in molecular dynamics simulations}
Classical molecular dynamics (MD) simulations were performed using the LAMMPS package \cite{THOMPSON2022108171}, to calculate the phonon, relaxing, system stability, and melting. All molecular dynamics calculations were implemented using two potentials: MLIPs and the classical Tersoff potential for SiC \textit{(Tersoff.SiC potential)} \cite{PhysRevB.39.5566}. The phonon dispersion in LAMMPS, we carried out by phonoLAMMPS code \cite{phonoLAMMPS}. The model penta silicene sheet contained 2400 atoms with $L_x$ = $L_y$ = 111.63 \text{\AA} and ATOMSK package \cite{HIREL2015212} produce the initial input model. The periodic boundary conditions are applied in the \textit{x} and \textit{y} Cartesian directions and in the \textit{z} direction, a fixed boundary with elastic reflection behavior is applied. The time step of the MD simulation was 1 fs. Relaxing and system stability of MD simulations are performed under canonical ensemble (NVT) and microcanonical ensemble (NVE). Melting are performed under canonical ensemble (NVT) and isothermal-isobaric ensemble (NPT). VESTA and OVITO softwares are used for 2D visualization of atomic configurations \cite{VESTA, ovito}.

\subsection{On-the-fly machine learning force field}
To efficiently construct an accurate machine learning interatomic potential (MLIP), an on-the-fly machine learning strategy was employed during the ab initio molecular dynamics (AIMD) simulations within VASP \cite{PhysRevB.54.11169, https://doi.org/10.1002/jcc.21057}. The AIMD simulations from 300 K to 700 K with a each step of 100 K are performed. The time steps used 1 fs, and the number for steps set up 10 000 time steps for each temperature model. The training data for each of these models were generated from canonical ensemble (NVT) simulations. The temperature control is achieved by the Nose-Hoover thermostat \cite{10.1063/1.449071, PhysRevA.31.1695}. During the training process, the weights for energy, forces, and stresses are assigned values 1, 0.1, and 0.001, respectively.

\section{Results and discussions}

\subsection{Structure and electronic properties}
Penta silicene is novel two-dimensional allotropes of silicon, which like penta graphene and the same symmetry  P42$_1$m (space group No.113) \cite{Shunhong}, show in \textcolor{blue}{[Fig. \ref{fig1s}(a) and \ref{fig1s}(b) in SI]}. Our phonon calculation result show that the penta silicene is unstable with space group No.113 and our result is in good agreement with previous report \cite{PhysRevApplied.11.064063}, show in \textcolor{blue}{[Fig. \ref{fig1s}(c), SI]}. \\The new penta silicene was generated by moving  Si$_3$ atoms. When one Si$_3$ atom moves towards and the other Si$_3$ atom moves away from the sheet, the penta silicene phase was generated by “tilting” and break the high symmetry of the system, it reduced the strong Coulomb repulsion, leading to stabilize the crystal structure. The symmetry of penta silicene is reduced from P42$_1$m (space group No.113) to P2$_1$ (space group No.4). Result above reported by Yaguang Guo et al (2019) \cite{PhysRevApplied.11.064063}. From the top view of Fig. \ref{fig1}(a) show that, the new phase is generated entirely of silicon pentagons, there are six silicon atoms in the primitive cell as denoted by red dashed lines. From the side view a tilting h= 0.185 \text{\AA} is observed and leading to a 2D sheet with a total thickness of d = 2.31 \text{\AA}. The lattice constants is a = b= 5.58\text{ \AA} respectively, our results is in good agreement with previous report \cite{doi:10.1021/acsami.9b21076}. To convenience the discussion process, we label the system penta silicene as group the sp$^3$- and sp$^2$-hybridized Si atoms as Si$_1$ and Si$_2$, respectively. The bond lengths show Si$_1$-Si$_2$ and Si$_2$-Si$_2$ are 2.39 and 2.28 \text{\AA}, both showing the double bond character. Specifically in penta silicene sheet, the bond angles $\theta_{\mathrm{Si_1\text{-}Si_2\text{-}Si_1}}$, and $\theta_{\mathrm{Si_1\text{-}Si_2\text{-}Si_2}}$ are $108.68^{\circ}$, and $100.25^{\circ}$, respectively.    
\\ Fig. \ref{fig1}(b) shows the first Brillouin zone with the high-symmetry k points labeled: $\Gamma$(0 0 0), X(0 1/2 0), S(1/2 1/2 0), and Y(1/2 0 0), respectively. We performed phonon dispersion calculations along high-symmetry lines in the first Brillouin zone to assess the experimental viability of penta silicene. Although a negligible imaginary frequency is observed near the $\Gamma$ point ($\Gamma$-right) over the entire frequency range,  the overall spectrum confirms the dynamical stability of the penta silicene. The acoustic branches exhibit linear dispersion near the $\Gamma$ point while the optical branches extend up to approximately 15 THz, show in Fig \ref{fig1}(c). The electronic band structure calculated that the penta silicene is an indirect semiconductor, with the valence band maximum located near the X-S points and the conduction band minimum appearing at the S-Y point and resulting in an indirect energy gap of 0.24 eV, consistent with previous research results \cite{doi:10.1021/acsami.9b21076}, show in Fig. \ref{fig1}(d).

\subsection{Thermodynamic properties}
\subsubsection{Training data for MLIPs}
We started generating data from AIMD calculations, \textcolor{blue}{[Fig. \ref{fig2s}, SI]} describes the structure configurations after 0.5 ps of AIMD at different temperatures. At low temperatures from 100 K to 400 K, penta silicene maintains a stable geometric structure with characteristic 5-rings. As the temperature gradually increases from 500 K to 900 K, under the influence of temperature, the silicon atoms vibrate strongly, leading to deformation in the structure. Within the simulated timescale of 0.5 ps, the penta silicene structure remains thermally stable up to 700 K. The choice of the dataset steps determines the convergence and predictive accuracy of the model. Furthermore, choosing the data from 100 K to 900 K allows the data to cover a wide range of the states, from near-equilibrum configuration to strongly distorted configurations at low to high temperatures.
\\Evaluating the performance of MLIPs in predicting energies and atomic forces under different thermodynamic conditions is essential. We assess the impact of AIMD simulation to generate training data by comparing the energy ($E_{\mathrm{RMSE}}$) and force ($F_{\mathrm{RMSE}}$) of root mean square  error (RMSE). To ensure consistency and eliminate size effects, all evaluations are performed using the same system containing 96 atoms. Our hypothesis is that employing AIMD simulations at different temperature points, while neglecting finite-size effects, to clarify the role of temperature in expanding the explored phase space. Within the same simulation time of 0.5 ps, AIMD simulations performed at different temperatures generate more diverse training data in terms of energies and atomic forces, thereby providing the richer training datasets. Consequently, comparing models trained at different temperatures for the same system size allows for the comprehensive evaluation of the MLIPs under different temperature regimes, respectively.
\\ Fig. \ref{fig3s}(a) \textcolor{blue}{[in SI]} show that the data training process at the temperature of 300 K with 10$^6$ training steps, showing the convergence of the DeepMD model through the variation of the RMSE calculated: the entire dataset (RMSE\_ALL), the energy (RMSE\_Energy), the force (RMSE\_Force), and especially also showing the learning curve of the training model, respectively. The blue line represents the training dataset and the orange line represents the validation dataset. The RMSE curves representing the entire data set, energy, and force all start at high values, then decrease sharply and stabilize after approximately 0.6 × 10$^6$ steps. The training and validation curves nearly overlap throughout the training process, indicating the absence of overfitting. The learning rate decays exponentially and becomes stable after the 0.4 × 10$^6$ steps, suggesting that the neural network weights achieve their optimal accuracy in the later stages of training beyond 0.4 × 10$^6$ steps. Fig. \ref{fig3s} (b) and (c) \textcolor{blue}{[in SI]} represent the error between the DFT values and the DeepMD model's predicted values through RMSE evaluation, including RMSE\_Energy and RMSE\_Force. These errors reflect the accuracy of the machine learning force field constructed on the test dataset.
\begin{table}[h]
	\centering
	\caption{Testing accuracy of the MLIP models for penta silicene at different temperatures by RMSE.}
	\label{tab:testing_accuracy}
	\begin{tabular}{c c c}
		\multicolumn{3}{c}{\textbf{Penta-silicene (96 atoms)}} \\
		$T$ (K) & $E_{\mathrm{RMSE}}$ (eV/atom) & $F_{\mathrm{RMSE}}$ (eV/\AA) \\
		100 & 6.10 $\times 10^{-5}$ & 5.97 $\times 10^{-3}$ \\
		200 & 3.00 $\times 10^{-5}$ & 5.95 $\times 10^{-3}$ \\
		300 & 8.09 $\times 10^{-6}$ & 4.16 $\times 10^{-3}$ \\
		400 & 7.39 $\times 10^{-6}$ & 4.26 $\times 10^{-3}$ \\
		500 & 1.09 $\times 10^{-5}$ & 4.29 $\times 10^{-3}$ \\
		600 & 1.88 $\times 10^{-5}$ & 6.74 $\times 10^{-3}$ \\
		700 & 1.05 $\times 10^{-5}$ & 5.37 $\times 10^{-3}$ \\
		800 & 1.00 $\times 10^{-5}$ & 5.48 $\times 10^{-3}$ \\
		900 & 2.64 $\times 10^{-5}$ & 1.06 $\times 10^{-2}$ \\
\label{t1}
	\end{tabular}

	\vspace{2mm}
	\footnotesize
	\raggedright
	The errors are reported for energy per atom ($E_{\mathrm{RMSE}}$) and total force components ($F_{\mathrm{RMSE}}$).
\end{table}
Table \ref{t1} presents the RMSE values for energy per atom and the average atomic force of the MLIPs at various temperatures for the penta silicene model. For all MLIPs the energy RMSE values fall within the range of  $10^{-6}$ to $10^{-5}$ eV/atom, while the force RMSE values range from $10^{-3}$ to $10^{-2}$ eV/Å, indicating that the MLIPs exhibit very high predictive accuracy. The largest RMSE for energy ($6.10 \times 10^{-5}$ eV/atom) and force ($5.97 \times 10^{-3}$ eV/\AA) were observed at 100 K, while the smallest errors were obtained at 300 K and 400 K. Although the penta silicene system exhibits high structural stability at 100 K, phase space sampling is limited due to poor dataset of thermal oscillations, resulting in a less diverse training dataset, thereby increasing prediction errors and reducing model accuracy. Conversely, at 300 K and 400 K, under the influence of higher temperatures, the penta silicene structure maintains a stable 5-ring, while exhibiting diverse thermal oscillations. This allows the DeepMD model to train a rich dataset of configurations and improves the accuracy in predicting atomic energy and interaction forces.\\
As the temperature increases from 500 K to 900 K, the RMSE of both energy and force tends to increase gradually, this reflects the increase in anharmonic vibrations in penta silicene under the influence of temperature, and the dataset also expands with a variety of convergent configurations. In the temperature range from 500 K to 800 K, the MLIPs still maintain high accuracy with $E_{\mathrm{RMSE}}$ and $F_{\mathrm{RMSE}}$ values. At 900 K, the RMSE increases significantly with $E_{\mathrm{RMSE}}$ = $2.64 \times 10^{-5}$ and $F_{\mathrm{RMSE}}$ = $1.06 \times 10^{-2}$, under the high temperature distorts the structure and degrades the crystal order of the penta silicene system, causing data interference during DeepMD model training and the process accurate description of atomic interactions in the system. The model's extremely high correlation coefficient of 0.99 provides further evidence of the very high accuracy and fit between the predicted physical quantities and the initial ab initio (DFT) calculations. Furthermore, the low value of $E_{\mathrm{RMSE}}$ and $F_{\mathrm{RMSE}}$ are essential for ensuring the stability and kinematic accuracy of subsequent MD simulation processes.

\subsubsection{Phonon dispersion}
In the previous section, we discussed the DeepMD model using RMSE indices for energy and force. To verify the ability of MLIP to describe the molecular dynamics properties of the penta silicene, we performed phonon spectral calculations using the LAMMPS package via PhonoLAMMPS \cite{phonoLAMMPS}. The results obtained were directly compared with calculations from DFT, Tersoff.SiC potential, and MLIP, thereby accurately assessing the potential of MLIP in molecular dynamics calculations. \\
\begin{figure}[h]
	\centering
	\includegraphics[scale=0.7]{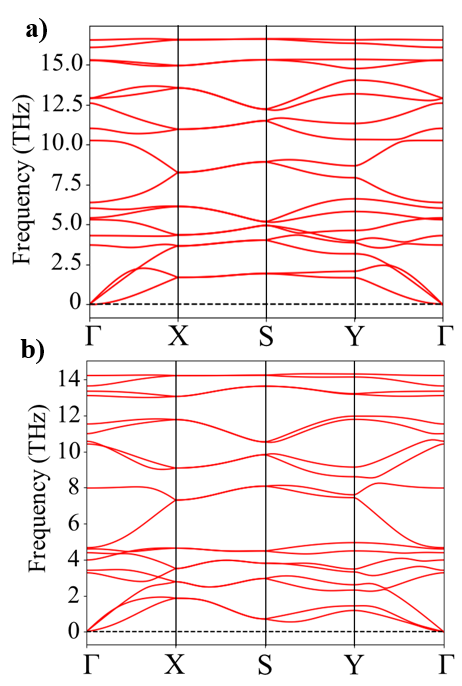}
	\caption{The phononLAMMPS calculations by two potentials: a) Tersoff.SiC potential. b) MLIP with a 4x4x1 supercel.}
	\label{fig2}
\end{figure}
\begin{figure*}[h]
	\centering
	\includegraphics[scale=0.85]{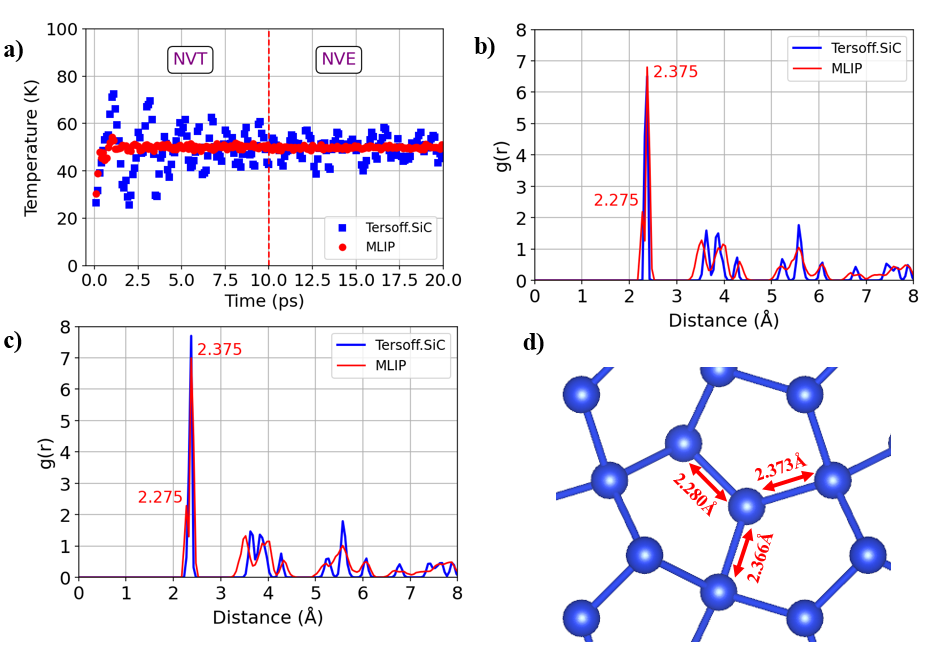}
	\caption{a) System temperature under NVT and NVE ensembles. b) Radial distribution function at 10 ps. c) Radial distribution function at 20 ps. d) Bond length between sp$^2$-sp$^2$ and sp$^2$-sp$^3$ hybridization.}
	\label{fig3}
\end{figure*}
The results of calculating the phonon plots in  Fig. \ref{fig2}(a), and (b) with 4x4x1 supercels of the two force fields MLIP and Tersoff.SiC potential show that the phonon spectra do not exhibit negative frequency branches (imaginary frequency) in the central Brillouin zone ($\Gamma$), while the DFT results show a very small negative value in Gamma (right corner) show in  Fig. \ref{fig1}(c). The penta silicene structure is completely dynamically stable as described by MLIP and Tersoff.SiC potential. The results of calculations with 2×2×1 and 3×3×1 supercells show a clear similar between MLIP and DFT \textcolor{blue}{[Fig. \ref{fig4s}, SI]}. Around the $\Gamma$ position, a small negative value exists, but this value decreases as the supercell size is increased; both potential models describe the phonon spectrum very well. The optical branches have frequencies below 15 THz. For the phonon spectrum from MLIP, only a small negative value appears at the acoustic branch around $\Gamma$ (right corner). In contrast, the phonon spectrum from the Tersoff.SiC potential shows a significant discrepancy; phonons in all supercells exhibit dynamic stability, and the values of frequencies are higher than 15 THz. This difference suggests that classical potentials are insufficient to accurately describe the intermolecular interactions in the penta silicene structure. Through the phonon calculation process, MLIP accurately reproduces both the low-frequency acoustic branches and the high-frequency optical branches, closely matching the results of DFT calculations.
\subsubsection{Relaxation and system stability}
To verify the optimal process and structural stability of penta silicene with 2400 atoms at 50 K \textcolor{blue}{[Fig. \ref{fig5s}, SI]}, we performed MD simulations. Fig. \ref{fig3}(a) shows the temperature variation over timesteps using two different interaction potentials: MLIP (red circles) and Tersoff.SiC potential (blue squares).
In the first 10 ps, penta silicene structure was optimized by NVT ensemble. Both models start at 30 K, then gradually increase the temperature and quickly converge to around 50 K, where the system reaches thermal equilibrium. Notably, after about 1.5 ps, the MLIP stabilized completely and maintained a temperature around 50 K with very small fluctuations, indicating that the machine learning potential accurately describes the interactions between atoms, approaching behavior at the quantum mechanical level. In contrast, with the Tersoff.SiC potential, the temperature fluctuated sharply throughout the first 10 ps, with large amplitudes in the range of 30–75 K. After 10 ps, penta silicene moved to the NVE ensemble to check system stability. Both models maintained stability around 50 K and reached thermal equilibrium before transitioning to the NVE ensemble. However, MLIP continued to maintain ideal stability with low thermal oscillation amplitudes, while the Tersoff.SiC potential exhibited consistently higher thermal oscillations. This phenomenon reflects the differences and limitations of  the Tersoff.SiC potential's in describing the structural stabilization and energy distribution of the system. 
\\Next, the radial distribution function g(r) provides insight into the atomic order and characteristic bond lengths in the penta silicene lattice. Fig. \ref{fig3}(b) shows the radial distribution function at 10 ps (the end of the structural optimization process) and Fig. \ref{fig3}(c) shows the end of the structural stability check. Both graphs show the characteristic sharp peaks of the structure in the short-distance (r $<$ 2.5 Å), where these peaks represent the closest bonds between atoms. Both force fields accurately determine the peak at 2.375 \AA, corresponding to the bond length, and the high intensity in the penta silicene structure. For MLIP, in addition to the main peak at 2.375 \AA, there is also a peak at 2.275 \AA, representing the shortest bond distance in the penta silicene structure. This result is consistent with the bond lengths between silicon atoms with sp$^2$–sp$^2$ hybridization, as verified and shown in Fig. \ref{fig3}(d). The second peak at 2.375 \AA, corresponds to the most common bond length of silicon atoms in 5-rings. Furthermore, sp$^2$–sp$^3$ bonds with lengths of 2.366 and 2.373 \AA \text{} are clearly reflected in the radial distribution function graph through the peak at 2.375 \AA. Conversely, for the  the Tersoff.SiC potential, the figure shows only a single peak at 2.375 \AA, corresponding to the bond length (sp$^2$–sp$^3$ hybridization) in the penta silicene structure. This indicates that MLIP fully and accurately describes the bond length distribution, as well as the various hybridization characteristics in the penta silicene structure.
\subsubsection{The phase transition of penta silicene}
After optimizing the structure at 50 K with 10$^4$ MD steps to achieve thermal equilibrium, the penta silicene model was heated from 50 K to 1000 K using MLIP and from 50 K to 2000 K using the Tersoff.SiC potential, with heating rates of $\gamma$ = 47.5 K/ps and $\gamma$ = 0.25 K/ps, respectively. The phase transition (T$_g$) was analyzed through the dependence of the potential energy per atom on the phase transition (T$_g$) temperature with MLIP; results with  the Tersoff.SiC potential show in \textcolor{blue}{[Fig. \ref{fig6s}(a), SI]}). In the low temperature region T $<$ T$_g$, the potential energy varies linearly with temperature, indicating that the system maintains structural order and thermal stability. When the temperature exceeds  T $>$ T$_g$, the potential energy deviates from the linear trend, forming a characteristic break point, which determines the melting temperature of penta silicene.
\begin{figure}[h]
	\centering
	\includegraphics[scale=0.57]{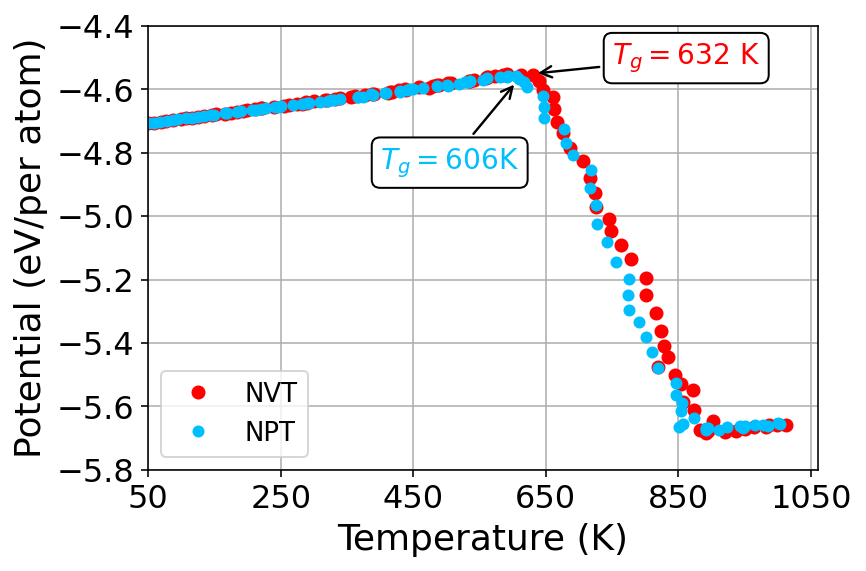}
	\caption{The phase transition (T$_g$) of penta silicene with MLIP.}
	\label{fig4}
\end{figure}
\\Fig. \ref{fig4} and \textcolor{blue}{[Fig. \ref{fig9s}(a), SI]} present the dependence of the energy per atom on the melting of penta silicene using MLIP and Tersoff.SiC potential in NVT (red circle) and NPT (deepskyblue circle) ensembles. 
\\With MLIP, in the low temperature range from 50 K to nearly 600 K (T $<$ $T_g$), the energy increases almost linearly with temperature for both ensembles from -4.71 eV/per atom at 50 K to -4.56 eV/per atom at 600 K. From the results, show that the difference between NVT and NPT ensembles were negligible, and the penta silicene structure maintains its stability. As the temperature increases and approaches the phase transition region, the total energy per atom begins to deviate from the linear path due to a sharp decrease, and a break point appears, which is the melting point of penta silicene. Specifically, NPT ensemble determined at T$_g$ = 606 K with the energy is -4.56 eV/per atom, and NVT ensemble at T$_g$ = 632 K with the energy is -4.57 eV/per atom, this difference demonstrates the role of boundary conditions of pressure and volume in influencing the stability and melting point of penta silicene. At high temperatures (after the phase transition point up to 850K), the energy values decrease sharply, indicating a significant drop in total energy and a strong structural transformation phase. From T $>$ 850 K, the potential energy approaches saturation, at an energy level of approximately -5.7 eV/per atom, which shows that the system has reached a stable configuration state after the melting point.\\
Fig. \ref{fig6s}(a) and (b) in \textcolor{blue}{SI} respectively present the results of the radial distribution function (g(r)) of penta silicene at two NVT and NPT ensembles with different temperature points. At 50 K, penta silicene is a stable (results have been presented in section 3.2.3). When the temperature increases to 150 K, approaching T $<$ T$_g$, the first peak position at 2.275 \AA \text{} is lost due to thermal vibration, but the peak position at 2.375 \AA \text{} still shows the Si-Si bond length. When the temperature is increased to the melting point, at approximately 632 K for NVT and 606 K for NPT, under the influence of temperature, the peaks begin to decrease and spread out, a sign of gradual disorder in the crystal lattice. Note that, we also carried out AIMD simulations at a large time step of 10 ps to check the thermal stability and present in the section 3.3. Fig. \ref{fig7s} and \ref{fig8s} in \textcolor{blue}{SI} show the penta silicene sheet exhibiting defects at the melting point. At high temperatures (750 K to 1000 K), g(r) completely lost its peak structure and weak oscillations around approximately 1 appeared, characteristic of the amorphous liquid state, consistent with the abrupt change in potential energy in the previous graph.\\
With Tersoff.SiC potential \textcolor{blue}{[Fig. \ref{fig9s}(a), SI]}, the potential energy versus temperature plot shows a distinct slope change, presenting the energy per atom of -3.88 eV/per atom at 50 k and -3.44 eV/per atom at 1600 K. Under NVT ensemble (red circle), $T_g$ = 1850 K with an energy of -3.39 eV/per atom, and NPT ensemble (deepskyblue circle), $T_g$ = 1780 K with the energy of -3.41 eV/per atom. The Tersoff.SiC potential's $T_g$ has an unusually high value when describing the melting point of penta silicene. Because  the Tersoff.SiC potential is a bond-order potential constructed to describe the oriented covalent bonds in group IV and the overstability of the tetrahedral lattice, and predicts higher melting temperatures than experimentally and overestimates the stability of the solid phase. This result has also been published by Stephen J. Cook \cite{PhysRevB.47.7686}.\\
The g(r) at low temperature 50 K, under both NVT and NPT ensembles \textcolor{blue}{[Fig. \ref{fig9s}(b) and (c), in SI]}, exhibits sharp, narrow, and distinct peaks with value g(r) = $2.375$ \AA, which reflects the order of the structure. As the temperature gradually increases towards the phase transition region, the intensity of the peaks g(r) decreases and their width increases due to the effect of temperature. At the melting point under NVT ensemble, the intensity of the radial distribution function is lower than under NPT ensemble. When observing the penta silicene sheet. \textcolor{blue}{[Fig. \ref{fig10s} and \ref{fig11s}, SI]}, under NPT ensemble, the disorder of the penta silicene sheet is greater than under NVT ensemble. As the temperature increases beyond the melting point, g(r) reflects the gradual disorder of the penta silicene structure, indicating a transition to an amorphous state.

\subsection{On-the-fly machine learning force fields}
To more accurately assess the thermal stability of the structure with large simulation time steps, the thermal stability of penta silicene was verified through ab initio molecular dynamics (AIMD) calculations incorporating machine learning force fields (MLFF). AIMD simulations were performed over a time interval of 10 ps at temperatures ranging from 300 K to 700 K \textcolor{blue}{[Fig. \ref{fig12s}, SI]}. The variation of total energy with the number of time steps is represented by the blue line, while the instantaneous temperature is represented by the red line.
\begin{figure}[h]
	\centering
	\includegraphics[scale=0.50]{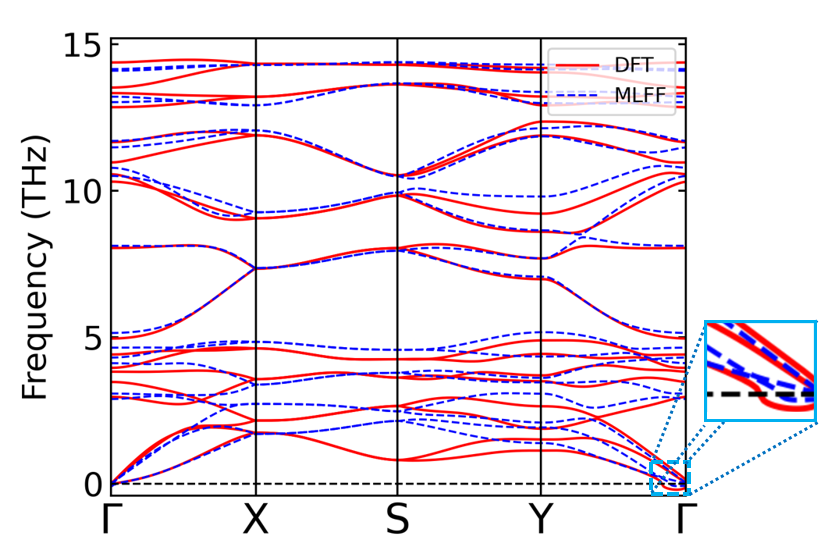}
	\caption{The compared phonon dispersion DFT and MLFF.  }
	\label{fig5}
\end{figure}
The results show that at temperatures of 300 K and 400 K, the total energy fluctuates slightly and no structural defects appear, indicating that penta silicene maintains good thermal stability. As the temperature increases to 500 K and 600 K, under the influence of heat, the amplitude of atomic vibrations increases significantly, leading to a corresponding increase in the amplitude of energy vibrations. However, the penta silicene structure remained stable at these two temperature points. Notably, at 700 K, penta silicene was strongly affected by the high temperature, leading to the appearance of structural defects during the simulation. With a large time step (10 ps), these deformations became more pronounced, indicating that the structure began to lose thermal stability. This result is consistent with the melting process of the structure, in which the phase transition temperature (T$_g$) was determined to be around 632 K, thereby further reinforcing the consistency of previous analyses in the melting of the structure and the MLIP model training data. Both phonon spectrum calculations using DFT (red line) and MLFF (dashed blue line) show that the phonon spectrum has frequencies below 15 THz. Particularly at Gama (right corner), both models clearly show a very small negative value (enlarged image). Although discrepancies still exist between the DFT and MLFF results at low-frequency branches, these results confirm that the penta silicene structure was dynamically stable and has high potential and accuracy between the two calculation methods MLFF and DFT.

\section{Conclusions}
\noindent  In summary, through ab-initio MD calculations, we constructed machine learning-based interaction potentials (MLIPs) using the DeepMD package. These potentials were trained to replicate the quantum mechanical accuracy of density functional theory (DFT) while maintaining significantly lower computational costs compared to pure AIMD. Training data were collected from AIMD simulations over a period of 0.5 ps, with very small energy and force errors, demonstrating the high training quality of the MLIP model.\\
From the constructed MLIPs, we demonstrated good correlation between the phonon spectra obtained from the MLIP simulations and the DFT results, and optimized and verified the structural stability of the system. Radial distribution function analysis revealed characteristic peaks at 2.275 and 2.375 Å, accurately reflecting the binding environment within the system, while the Tersoff.SiC potential only reproduced a peak at 2.375 Å. Furthermore, the firing of the penta silicene plate in the temperature range of 50 K to 1000 K clearly showed the phase transition and glass transition/melting temperature when using different ensembles, with T$_g$ = 632 K in NVT and T$_g$ = 606 K in NPT. These results were verified by AIMD simulations incorporating machine learning field force (MLFF) with a timesteps of 10 ps. The results clearly demonstrate the phase transition process of the penta silicene sheet within the LAMMPS simulation framework.\\
These results not only confirm the reliability of the DeepMD model for penta silicene but also show the potential for widespread application of machine learning potentials to other two-dimensional materials, opening up the possibility of performing large-scale, highly accurate simulations for many different nano material systems.

\section*{ACKNOWLEDGMENTS}
\noindent This research is funded by Vietnam National Foundation for Science and Technology Development (NAFOSTED)

\bibliography{bib} 
\bibliographystyle{apsrev}

\section*{Supporting Information}
Please refer to the supplementary information (SI) for details on the penta silicene model, AIMD, training data, phonon, radial distribution function, and the phase transition of Tersoff.SiC.

\begin{figure*}[ht!]
	\centering
	\includegraphics[scale=0.85]{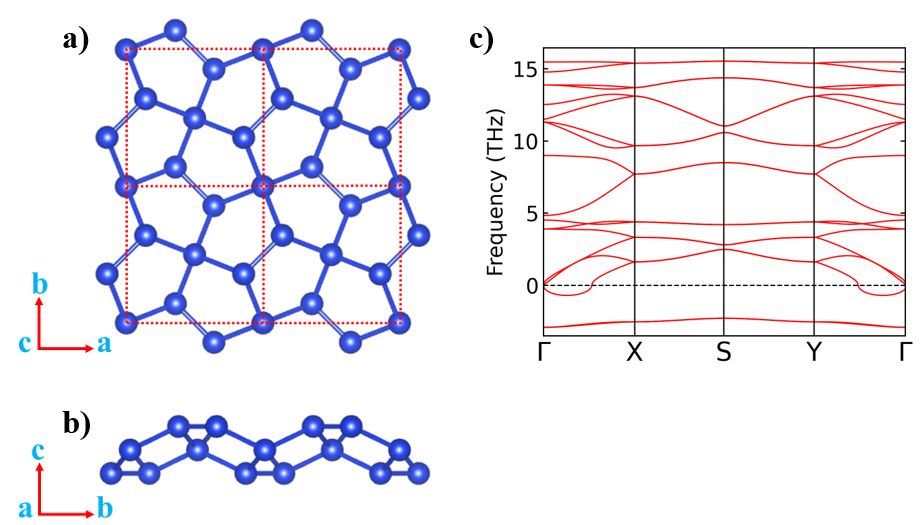}
	\caption{a) and  b) Top and  side view of the penta silicene is the symmetry P42$_1$m (space group No.113). c) The phonon dispersion are calculated with 2x2x1 supercell.}
	\label{fig1s}
\end{figure*}

\begin{figure*}[t]
	\centering
	\includegraphics[scale=1.3]{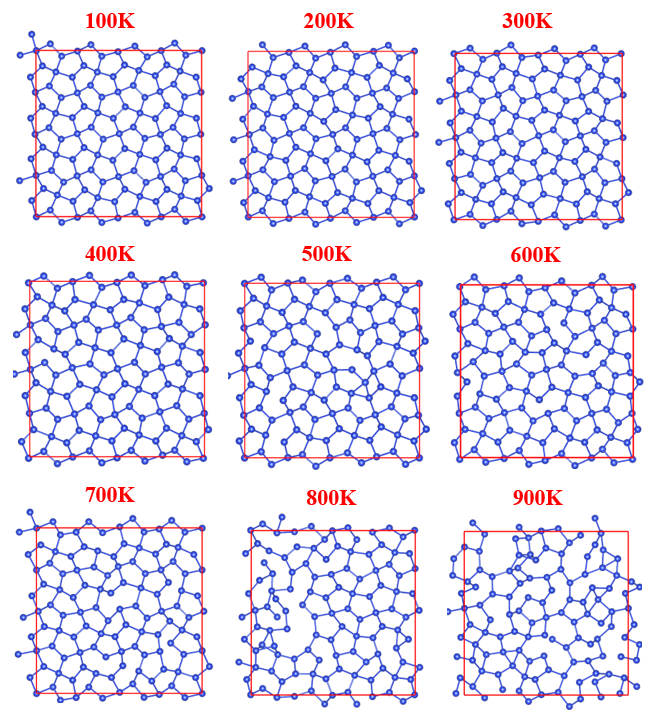}
	\caption{The AIMD simulations of penta silicene from 100K to 900K for the process generation data. The structures are the final configurations after AIMD simulations are done.}
	\label{fig2s}
\end{figure*}

\begin{figure*}[t]
	\centering
	\includegraphics[scale=0.75]{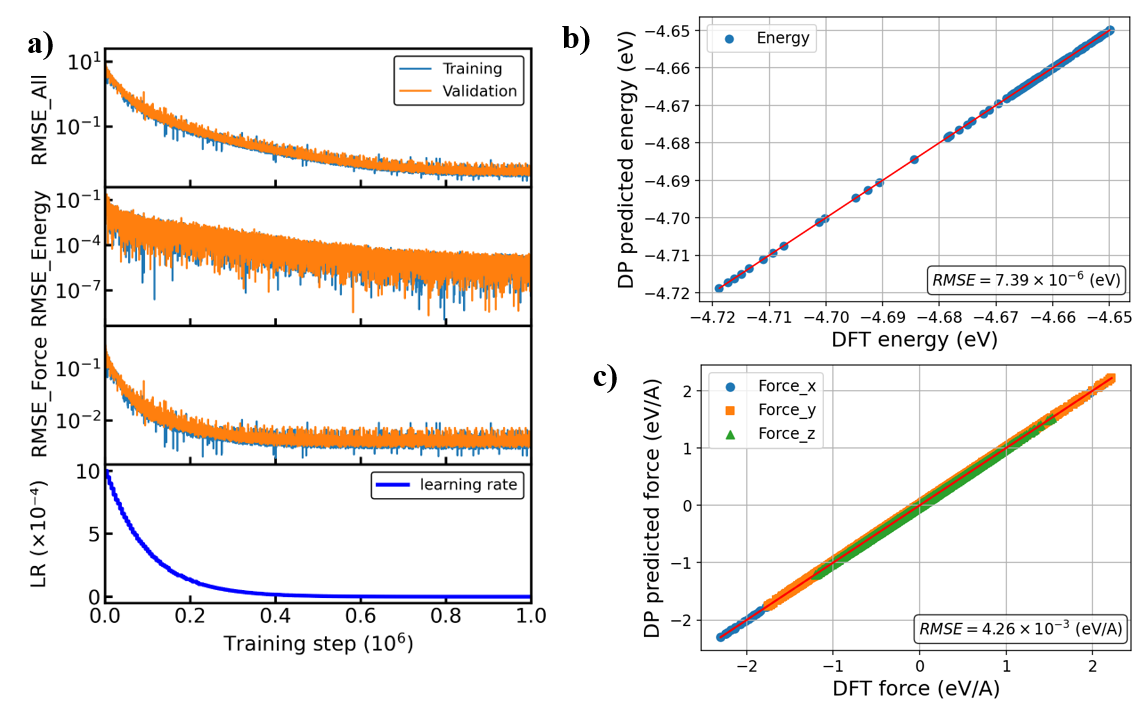}
	\caption{a) Training loss with steps at 300K. b) Test of the predictive accuracy of two DeepMD compared with DFT energies per atom. c) Test of the predictive accuracy of two DeepMD compared with DFT forces.}
	\label{fig3s}
\end{figure*}

\begin{figure*}[h]
	\centering
	\includegraphics[scale=0.95]{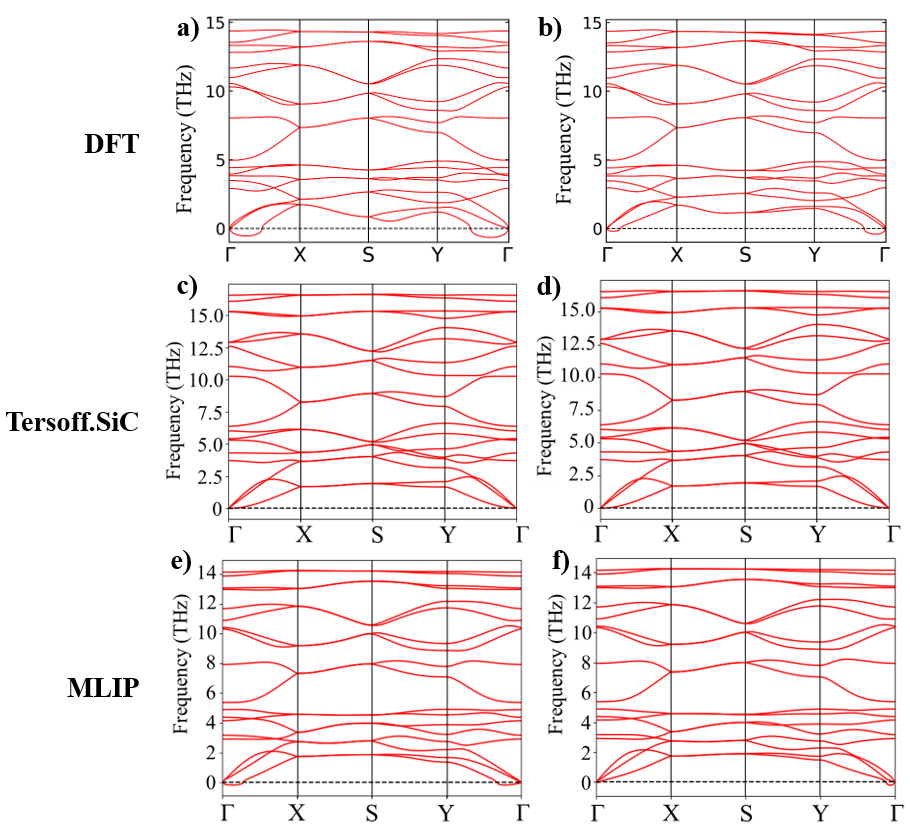}
	\caption{The phonon calculations by DFT, Tersoff.SiC potential, and MLIP: a, c, and e) a 2x2x1 supercell. b, d, and f) a 3x3x1 supercell.}
	\label{fig4s}
\end{figure*}

\begin{figure*}[h]
	\centering
	\includegraphics[scale=1.1]{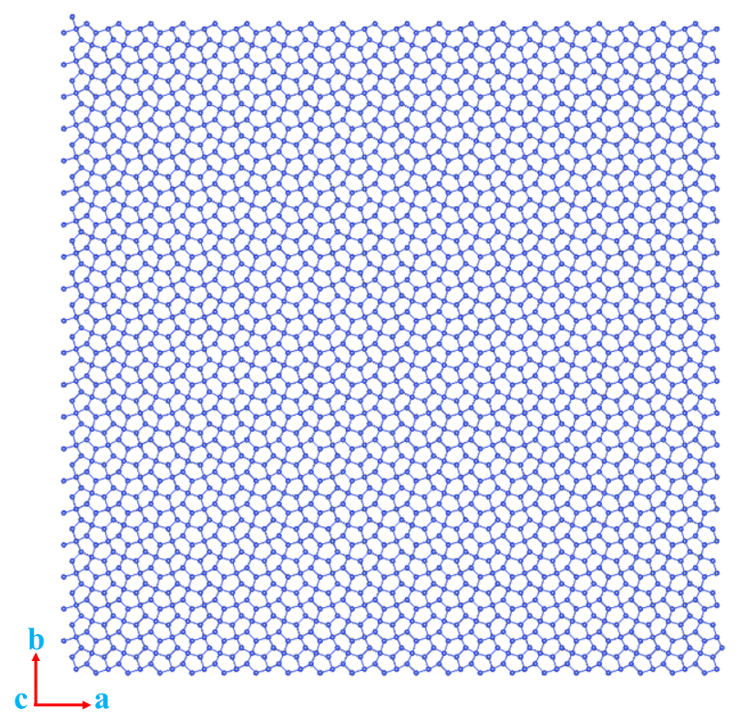}
	\caption{The initial penta silicene with 2400 atoms.}
	\label{fig5s}
\end{figure*}

\begin{figure*}[h]
	\centering
	\includegraphics[scale=1.0]{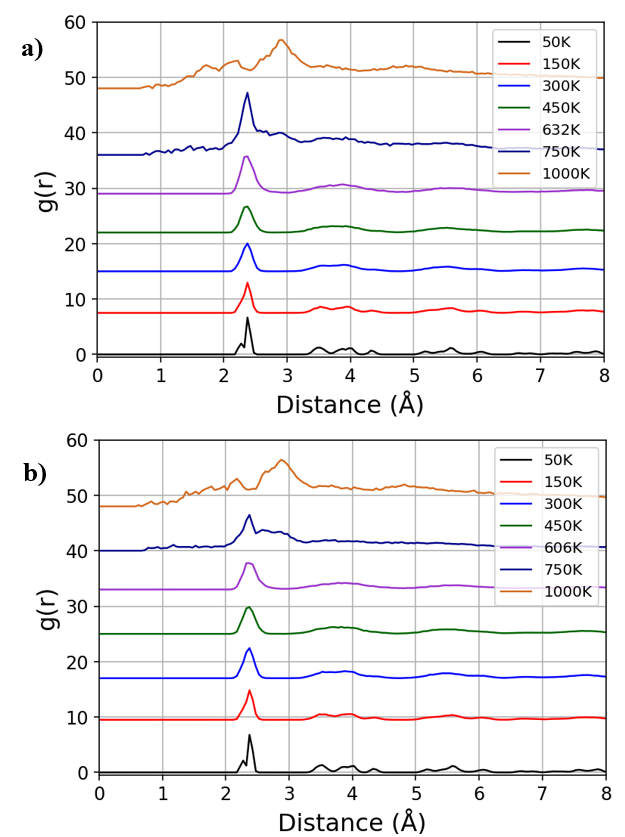}
	\caption{The radial distribution with MLIP. a) NVT ensemble. b) NPT ensemble.}
	\label{fig6s}
\end{figure*}

\begin{figure*}[h]
	\centering
	\includegraphics[scale=1.1]{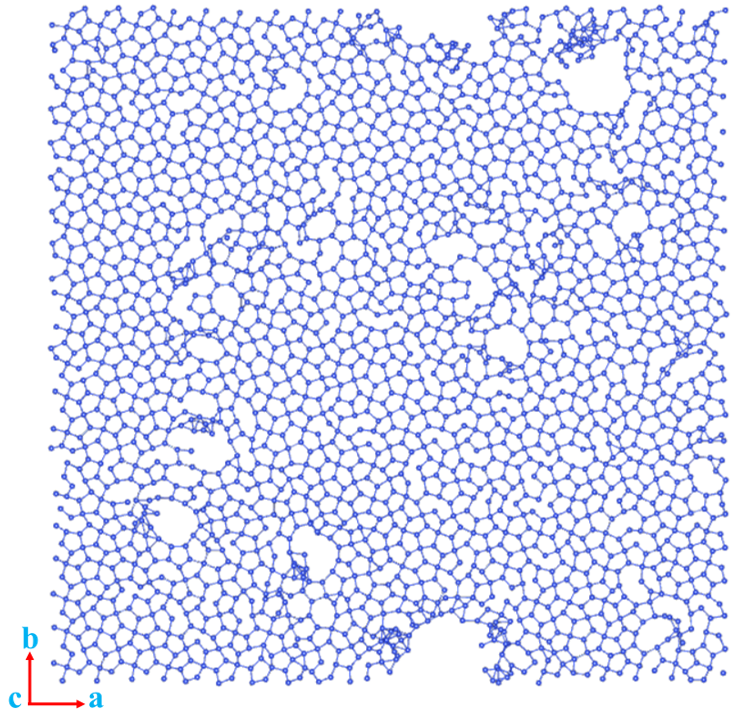}
	\caption{Configuration penta silicene at 632K with the NVT ensemble with MLIP.}
	\label{fig7s}
\end{figure*}

\begin{figure*}[h]
	\centering
	\includegraphics[scale=1.1]{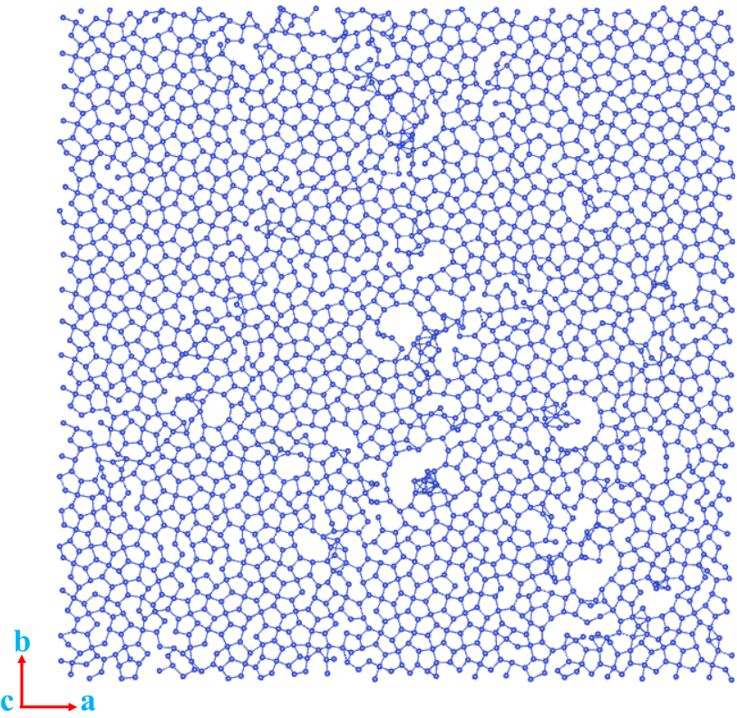}
	\caption{Configuration penta silicene at 606K with the NPT ensemble with MLIP.}
	\label{fig8s}
\end{figure*}

\begin{figure*}[h]
	\centering
	\includegraphics[scale=0.88]{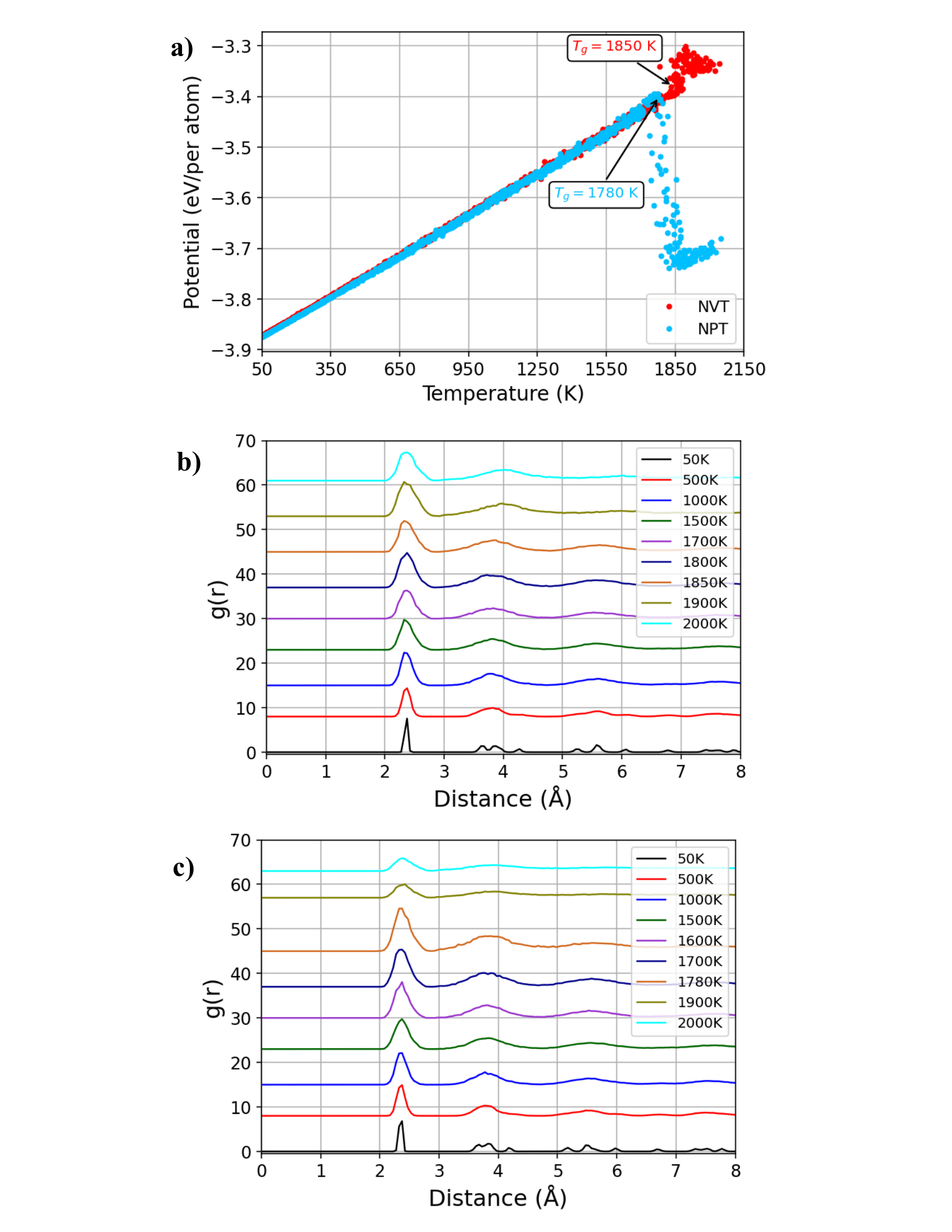}
	\caption{a) The phase transition (T$_g$) and the radial distribution of penta silicene with the Tersoff.SiC potential. b) NVT ensemble. c) NPT ensemble.}
	\label{fig9s}
\end{figure*}

\begin{figure*}[h]
	\centering
	\includegraphics[scale=1.1]{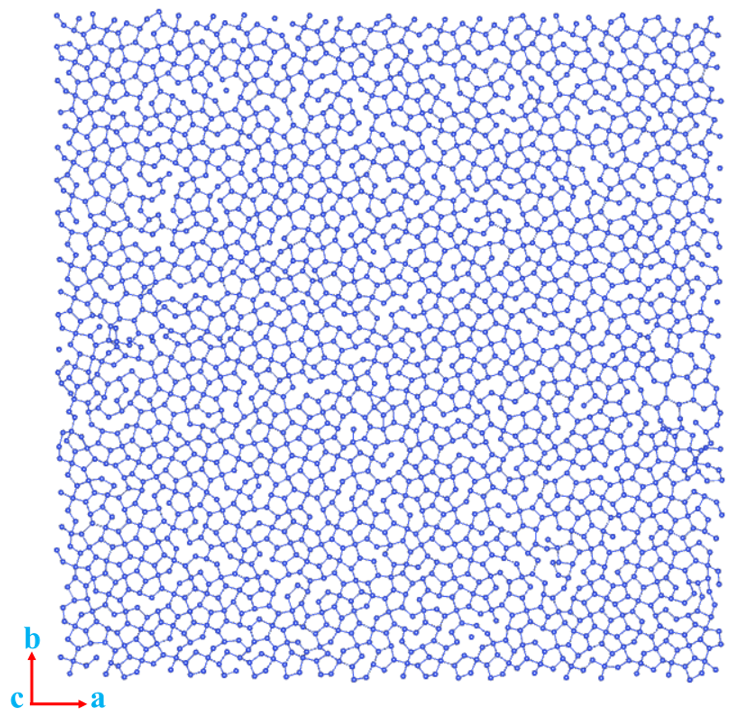}
	\caption{Configuration penta silicene at 1850K with the NVT ensemble with the Tersoff.SiC potential}.
	\label{fig10s}
\end{figure*}

\begin{figure*}[h]
	\centering
	\includegraphics[scale=1.1]{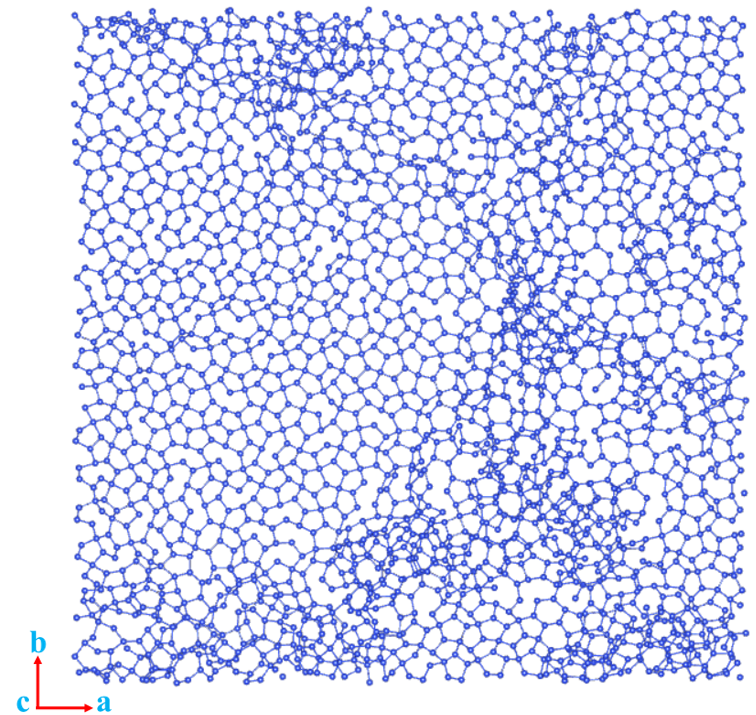}
	\caption{Configuration penta silicene at 1780K with the NPT ensemble with the Tersoff.SiC potential}.
	\label{fig11s}
\end{figure*}

\begin{figure*}[h]
	\centering
	\includegraphics[scale=0.85]{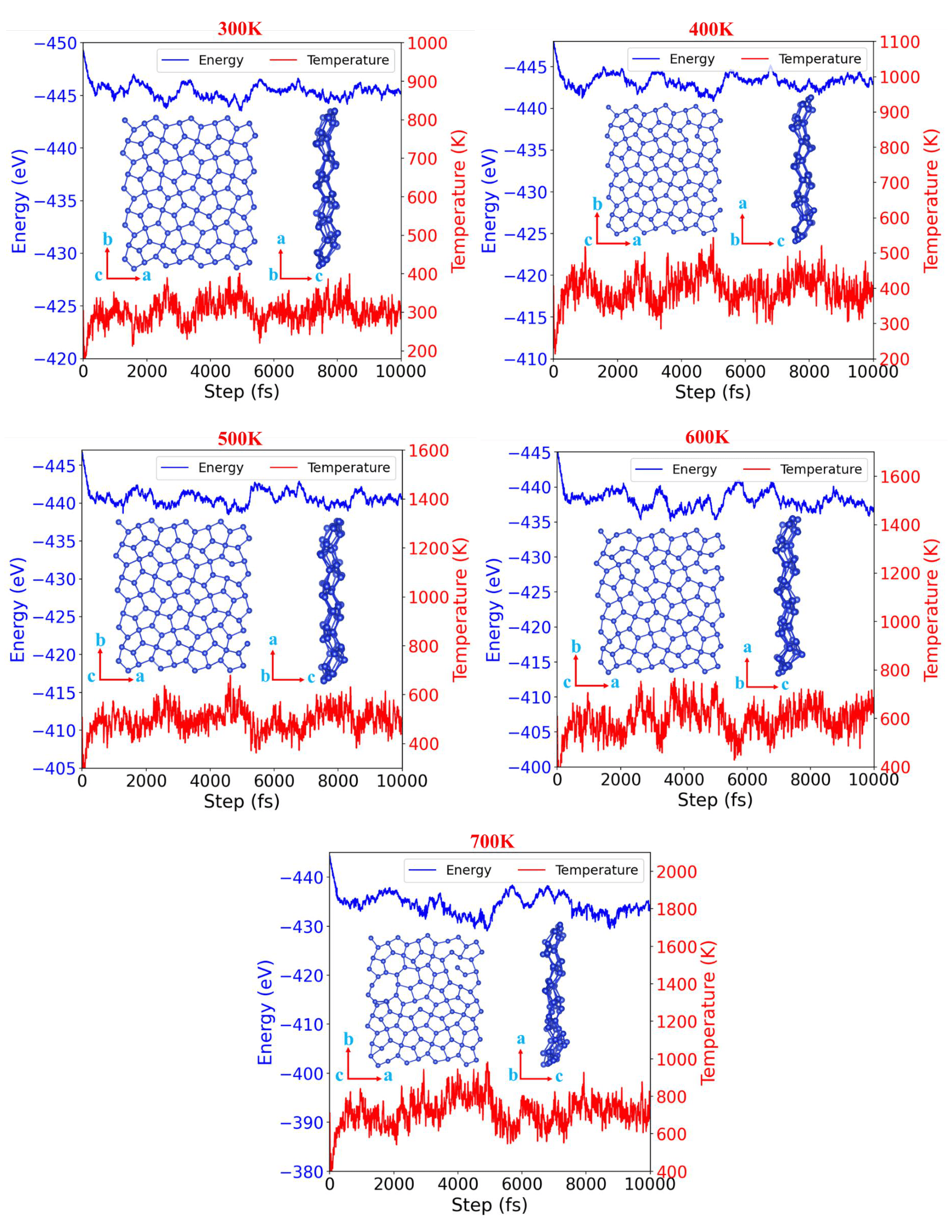}
	\caption{The AIMD simulations of penta silicene from 300 K and 700 K at 10 ps. The structures in the figure are the final configurations after AIMD simulations.}
	\label{fig12s}
\end{figure*}

\end{document}